\documentclass[ 
pra,
bibnotes,
 amsmath,amssymb,
 aps,
dvipdfmx, 
twocolumn
]{revtex4-2}

\usepackage{graphicx}
\usepackage{dcolumn}
\usepackage{bm}
\usepackage{braket}
\usepackage{cases}
\usepackage{amssymb}
\usepackage{amsmath}
\usepackage{color}
\usepackage{url}

\begin{document}

\title{Observing phase jumps of solitons in Bose--Einstein condensates}
 
\author{ Kazuma Ohi$^{}$, Shohei Watabe$^{}$, and Tetsuro Nikuni$^{}$}  
\affiliation{$^{}$ Department of Physics, 
Tokyo University of Science, Shinjuku, Tokyo 162-8601, Japan}

\begin{abstract} 
The phase difference of the macroscopic wave function is a unique structure of the soliton in an atomic Bose--Einstein condensate (BEC). However, experiments on ultracold atoms so far have observed the valley of the density profile to study the dynamics of solitons. 
We propose a method to observe the phase difference of a soliton in a BEC by using an interference technique with Raman and rf pulses. 
We introduce a phase jump factor, which is an indicator to measure the phase difference between two points. 
It is demonstrated by using the projected Gross--Pitaevskii equation that an interference density ratio, the density ratio of two-component BECs after the Raman and rf pulses, reproduces the phase jump factor well. This technique will become an alternative method to study the decay and breakdown of a phase imprinted soliton in atomic BECs. 
\end{abstract}
\maketitle


\section{introduction}

Soliton is ubiquitous, as seen in fluids, polyacetylene~\cite{Su1698}, plasma~\cite{Ikezi1970}, Jovian atmosphere~\cite{MAXWORTHY1976261}, as well as Bose--Einstein condensates (BECs) in ultracold atoms~\cite{Burger1999,Denschlag2000,Dutton2001}.
One of the interesting properties of solitons in BECs is the existence of the phase of macroscopic wave functions, which is completely different from other classical systems, where the dark soliton in a BEC shows the $\pi$-phase jump. 
Thanks to this degree of freedom, the soliton is indeed created by using the phase imprinting method~\cite{Denschlag2000}.

The decay and breakdown of the soliton in BECs have been studied extensively and intensively~\cite{Brand2002,Burger1999,Anderson2001,PhysRevLett.89.110401,PhysRevLett.113.065301, PhysRevLett.113.065302,PhysRevA.84.043640,Aycock:2017eu}, 
where the soliton is disrupted through the snake instability~\cite{Brand2002,Burger1999,Anderson2001,PhysRevLett.89.110401}, through 
scattering of Bogoliubov quasi-particles and thermal atoms~\cite{PhysRevA.84.043640}, or through externally imposed impurities~\cite{Aycock:2017eu}. 
The decay of a dark soliton at finite temperature in the highly-elongated geometry was studied by using the Zaremba--Nikuni--Griffin (ZNG) theory~\cite{Jackson2007}. 
This study included the finite-temperature effect by using the Boltzmann equation for the thermal cloud. 
These studies on the disruption of solitons focused on the time-dependence of the density profiles.

In the earlier study~\cite{Ohya:2019uf}, the decay of the phase-imprinted dark soliton in two-dimensional geometry at nonzero temperatures has been studied by focusing on the fidelity of the classical fields. 
However, it was later found that the fidelity cannot characterize the decay of solitons~\cite{Ohi_unpublished}. 
The scaling of the decay rate of the fidelity is actually the same with and without the dark soliton. Therefore, another measure of the decay and breakdown of the soliton is needed. 
Another problem in the earlier study~\cite{Ohya:2019uf} is that the decay of the soliton was discussed in terms of the system energy rather than the temperature.
This is due to the projected Gross--Pitaevksii equation (PGPE)~\cite{Davis_2001, Davis2001, Davis2002, Davis2003, Blair2005, Blakie2007, Blakie2008, Blair2008} for an isolated system, which is inconvenient for experiments.

Although the phase, which is a complement of the number of particles (density) of the macroscopic quantum systems, 
is an important quantity of the soliton in BECs, a method to observe the phase jump of the soliton has been lacking.
If we could observe the phase jump of solitons, it would provide an alternative way to study the decay and breakdown of solitons in BECs. 
In this study, we propose a method to observe the phase difference of a soliton in a BEC based on the technique to observe the supercurrent decay in an annular BEC~\cite{Moulder2012}. 
We introduce a phase jump factor to measure the phase difference between two points in a BEC. 
Using the (stochastic) PGPE, we demonstrate that the phase jump factor is well reproduced by the interference density ratio of a two-component BEC after applying Raman and rf pulses.
This method will become an alternative tool to study the decay and breakdown of solitons in atomic BECs.

\section{PGPE and SPGPE}

Our simulation for soliton dynamics is based on the PGPE~\cite{Davis_2001, Davis2001, Davis2002, Davis2003, Blair2005, Blakie2007, Blakie2008, Blair2008} in the two-dimensional system, where the bosonic field operator $\hat \psi$ is projected onto the coherent region ${\bf C}$ as $\hat \psi_{\bf C} = {\mathcal P}_{\bf C} [\hat \psi ]$ by using the projection operator ${\mathcal P}_{\bf C} [ \, \cdot \, ]$, and then replaced with the classical field $\psi_{\bf C}$. 
The dynamics of the classical field $\psi_{\bf C}$ is described by the following equation~\cite{Davis_2001,Blakie2008}: 
\begin{align}
i \hbar \frac{\partial \psi_{\bf C}({\bf r},t) }{\partial t} 
= 
{\mathcal P}_{\bf C} 
\left [
\hat H_{\bf C} \psi_{\bf C} ({\bf r},t)  
\right ],
\label{PGPE}
\end{align}
where $\hat H_{\bf C} \equiv \hat H_{0} + g | \psi_{\bf C} ({\bf r},t) |^2$ with  
the single-particle Hamiltonian $\hat H_{0} = - \frac{\hbar^2}{2m} \nabla^2 + U ({\bf r})$ with atomic mass $m$ and 
$g = 4 \pi \hbar^2 a / m $ is a coupling constant with the $s$-wave scattering length~\cite{PhysRevA.3.1067}. 
In this paper, we assume an ultracold Bose gas in a harmonic trap potential $U({\bf r}) = m (\omega_x^2 x^2 + \omega_y^2 y^2 + \omega_z^2 z^2) / 2$ where the $z$-axis degree of freedom is frozen out with $\omega_z \gg \omega_{x,y}$. 
This system is effectively described by a two-dimensional Bose gas with a two-dimensional harmonic trap and a dimensionless effective 2D coupling constant $\tilde g = g/(\sqrt{2\pi}a_z)$ with $a_z = \sqrt{\hbar /m \omega_z}$~\cite{PhysRevA.79.033626}. 
The critical temperature of the ideal Bose gas in the two-dimensional harmonic trap is given by 
$k_{\rm B} T_{\rm c} =   \sqrt{6\hbar^2 \omega_x \omega_y N_{\rm tot} / \pi^2} $, where $N_{\rm tot}$ is the total number of particles.

The projection operator onto the coherent region can be described in terms of the eigenfunction $u_\nu ({\bf r})$ of the single-particle Hamiltonian $\hat H_0$, which satisfy $\hat H_0 u_\nu ({\bf r}) = \epsilon_\nu u_\nu ({\bf r})$. Here, $u_\nu ({\bf r})$ are orthonormal eigenfunctions of the harmonic potential system and $\epsilon_\nu$ is its eigenvalue given by $\epsilon_{\nu} = \hbar \omega_x \left ( n_x + 1/2 \right ) + \hbar \omega_y \left ( n_y + 1/2 \right )$ with a set of quantum numbers $\nu = (n_x, n_y)$. 
The classical field in the coherent region ${\bf C}$ is then given by 
\begin{align}
    \psi_{\bf C} ({\bf r}, t) = \sum\limits_{\nu \in {\bf C}} c_\nu (t) u_{\nu} ({\bf r}) , 
    \label{titen}
\end{align}
where the number of particles in the coherent region is $N_{\bf C} = \sum\limits_{\nu \in {\bf C}} |c_\nu|^2$. 
By applying the form \eqref{titen} to Eq.~\eqref{PGPE}, the PGPE reduces to the equation of motion for coefficients 
\begin{align}
    i \hbar 
    \frac{d c_\nu (t)}{dt} = A_\nu (t) ,
\end{align}
where $A_\nu (t) \equiv   \epsilon_\nu c_\nu (t) + g F_\nu (t)$ with  
$F_\nu (t) \equiv \int d{\bf r} u_{\nu}^* ({\bf r}) |\psi_{\bf C} ({\bf r},t)|^2 \psi_{\bf C} ({\bf r},t)$. 

The PGPE describes the equation of the motion for the classical field in the coherent region. The system described by this equation can be regarded as an isolated system for the coherent region, where the interaction between the coherent and incoherent regions is neglected. 
Because of this isolation, controlling the temperature in the PGPE is not feasible in the simulation~\cite{Rugh2001}. 
The SPGPE overcomes this problem, where the interaction between the particles in the coherent region and the incoherent region (heat bath) is effectively included \cite{Blakie2008}:
\begin{align}
d\psi_{\bf C}({\bf r},t) 
    = & 
\frac{1}{\hbar}  {\mathcal P}_{\bf C} 
    [ -i\hat H_{\bf C} \psi_{\bf C} ({\bf r},t) dt
    \nonumber 
    \\ 
    & 
    + 
\gamma (\mu - \hat H_{\bf C}) \psi_{\bf C} ({\bf r},t) dt
    + 
     dW({\bf r}, t)
    ], 
    \label{SPGPE}
\end{align} 
where $\mu$ is the chemical potential of the heat bath. 
The constant $\gamma$ gives the thermal damping, given by 
\begin{align}
    \gamma \equiv 
    & \gamma_0 
    \sum\limits_{k=1}^\infty 
    \frac{e^{\beta \mu (k+1)}}{e^{2\beta \epsilon_{\rm cut} k}} 
    \Phi [e^{\beta (\mu - \epsilon_{\rm cut})} , 1, k]^2, 
\end{align}
where $\gamma_0 = 4 m k_{\rm B}T a^2 / (\pi \hbar^2)$, and $\Phi$ is the Lerch transcendent. 
Here, $k_{\rm B}$ is the Boltzmann constant, $T$ the temperature of the heat bath, and $\beta=1/k_{\rm B}T$ the inverse temperature. 
The first term in \eqref{SPGPE} is the same as the PGPE, and the second and third terms describe the effects of the interaction between particles in the coherent and incoherent regions. 
In particular, the second term in \eqref{SPGPE} brings the system into thermal equilibrium with the chemical potential $\mu$, where the damping rate associated with the equilibration is given by $\gamma$. 
The third term in \eqref{SPGPE} gives the noise from the thermal bath, where $dW$ satisfies the relations, 
\begin{align}
    \langle dW ({\bf r}, t) \rangle = & 0 , 
    \\
    \langle dW^* ({\bf r},t) dW ({\bf r}', t) \rangle = & 2 \gamma k_{\rm B} T 
    \hbar \delta_{\bf C} ({\bf r}, {\bf r}') dt,  
    \\
    \langle dW ({\bf r},t) dW ({\bf r}',t) \rangle = & 
    \langle dW^* ({\bf r},t) dW^* ({\bf r}',t) \rangle 
    = 0.
\end{align}
Here, $\delta_{\bf C} ({\bf r}, {\bf r}')$ is the delta-function in the coherent region defined by $\delta_{\bf C} ({\bf r}, {\bf r}') = \sum\limits_{\nu \in {\bf C}} \phi_\nu ({\bf r}) \phi_\nu^* ({\bf r})$. 

The SPGPE in \eqref{SPGPE} can also be reduced into the equation of motion for coefficients, given by 
\begin{align}
    d c_\nu (t) 
    = & 
         \frac{1}{\hbar} \{ - i A_\nu (t) dt + \gamma [\mu c_\nu (t) - A_\nu (t)] dt + dW_\nu (t) \}, 
\end{align} 
where $dW_\nu (t)$ is the noise for the state $\nu$ given by 
\begin{align}
    dW_\nu (t) \equiv \int d{\bf r} u_\nu^* ({\bf r}) dW ({\bf r}, t), 
\end{align}
which satisfies the relations  
\begin{align}
    \langle dW_\nu (t) \rangle = & 0, 
    \\ 
    \langle dW_\nu^* (t) dW_{\nu '} (t') \rangle = & 
    2 \gamma k_{\rm B}T \hbar \delta_{\nu,\nu'}, 
    \\ 
    \langle dW_\nu (t) dW_{\nu'} (t') \rangle = & 
    \langle dW_\nu^* (t) dW_{\nu'}^* (t') \rangle = 0. 
\end{align}
In contrast to the PGPE, the SPGPE involves the chemical potential and the temperature as control parameters.
The system after the long-time evolution in the SPGPE reaches the thermal equilibrium state with a given chemical potential and temperature. 
This fact is useful for controlling the initial condition of BECs at thermal equilibrium before imprinting the soliton. 
See also Appendix~\ref{method}. 

\begin{figure*}[tbp]
    \begin{center}
    \includegraphics[clip,width=13.0cm]{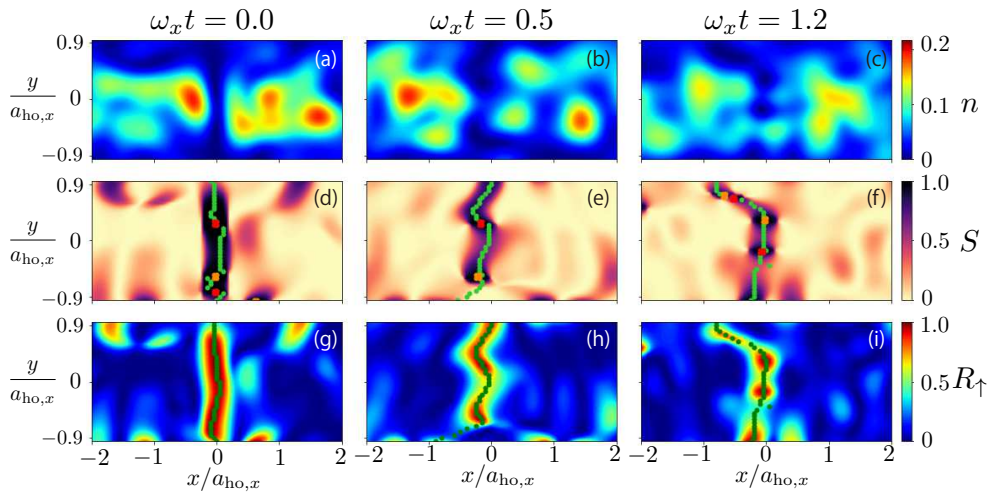}
    \end{center}
    \caption{
    Spatial profiles of the density $n$ (a)-(c), the phase jump factor $S$ (d)-(f), and the interference density ratio $R_{\uparrow}$ (g)-(i). 
    In (d)-(f), yellow (red) dots are the quantum vortices with (counter-)clockwise rotation, and the green line represents the soliton path $l_{S}$. 
    In (g)-(i), the green line represents the soliton path $l_{R_\uparrow}$. 
    We used $\omega_y / \omega_x = 5$, $T = 0.8T_{\rm c}$, $g = 0.1 \hbar^2/m$, and $N_{\rm tot} = 3000$. 
    }
    \label{p:hikabu}
    \end{figure*}

\section{Phase Jump Factor}

We apply the phase imprinting method~\cite{Denschlag2000} to the nonzero temperature BEC to create dark solitons.
Based on this method, we modify the phase $\theta (x,y)$ of the classical field in the stationary state to $\theta (x,y) + \pi$ for $x<0$ at $t=0$ . 
SPGPE provides this stationary state with a specified value of the chemical potential $\mu$ and temperature $T$. 
This phase imprinting generates the phase jump at $x=0$, and the dark soliton emerges at the same position (Fig.~\ref{p:hikabu}). (See also Appendix~\ref{AppendixB}.)
After the soliton emerges, we simulate the breakdown of the soliton at nonzero temperature by using PGPE. 
In contrast to the absolute zero case, the density distribution of the classical field at nonzero temperatures is significantly disturbed with time evolution due to thermal excitations. 

In this paper, we introduce a metric for studying the complement of the density, the phase jump factor $S(x,y)$, which characterizes the phase difference of the soliton at BECs:
\begin{align}
    S(x,y) \equiv \frac{1}{2} \{ 1 - \cos [\Delta \theta (x,y)]\}, 
\label{eqS}
\end{align}
where 
$\Delta \theta (x,y ) = \theta (x + d/2, y) - \theta (x - d/2, y)$. 
Here, $\theta (x,y)$ is the phase of the classical field. 
This gives the phase difference between two points at a distance $d$.
It is reasonable to take the healing length $d = \xi$ of BECs to study the soliton. 
The minimal phase difference $\Delta \theta = 0$, which means no phase difference, gives the minimal phase jump factor $S(x,y) = 0$. 
The maximal phase difference $\Delta \theta = \pm \pi$, which indicates the existence of the soliton, gives the maximal phase jump factor $S(x,y) = 1$ (Fig.~\ref{p:hikabu}). 
This phase jump factor clearly shows the existence of the phase jump, as well as the breakdown of the soliton into the vortex pairs.

\begin{figure*}[tbp]
\begin{center}
\includegraphics[clip,width=13.0cm]{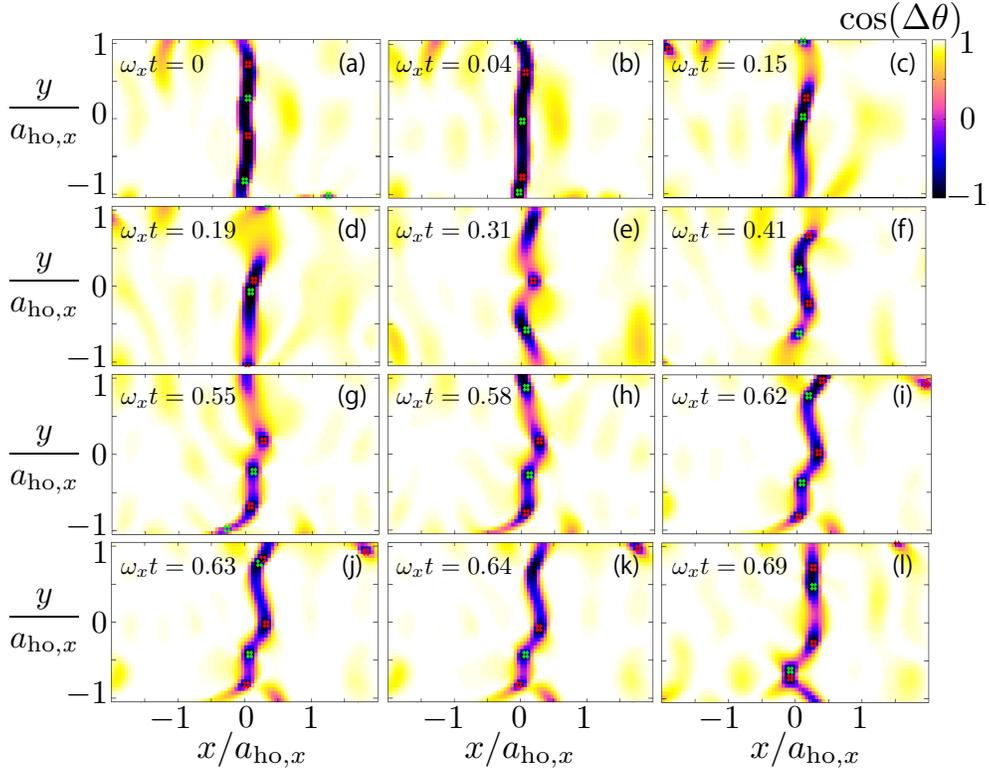}
\end{center}
\caption{
Time-evolution of the phase difference $\cos (\Delta \theta (x,y))$ at $T = 0.7T_{\rm c}$. 
We used the same parameters as in Fig.~\ref{p:hikabu}. 
Green and red dots represent the (ghost) vortex and anti-vortex, respectively. 
}
\label{p:phasedifference}
\end{figure*}

An interesting property of the sotlion breakdown can be shown in the time-dependence of the phase difference. The key is the pair of the vortex and antivortex as in Ref.~\cite{Gawryluk:2021wa}. 
First, immediately after the phase is imprinted, ghost vortices and anti-vortices are trapped alternately in the soliton (Fig.~\ref{p:phasedifference} (a)). We can find the region where the phase jump becomes small, when the nearest neighbor ghost vortex and anti-vortex move away (Fig.~\ref{p:phasedifference} (c)). This can become the break point of the soliton. 
We then call a string with a large phase difference a chopped soliton. 
A ghost vortex pair is trapped in the chopped soliton (Fig.~\ref{p:phasedifference} (d)), and when a ghost anti-vortex moves away from a paired ghost vortex, the ghost anti-vortex turns into a real anti-vortex (Fig.~\ref{p:phasedifference} (e)). Interestingly, because of the strong fluctuation, the reconnection happened among this real anti-vortex, the remaining ghost vortex, and another ghost vortex pair, which makes a longer chopped soliton (Figs.~\ref{p:phasedifference} (f)). The disconnection and connection are repeated (Figs.~\ref{p:phasedifference} (g), (h) and (i)). The ghost vortex pair can be annihilated in the chopped soliton (Figs.~\ref{p:phasedifference} (i), (j), and (k)). 

The phase difference $\Delta\theta$ is not observed directly in experiments; however, 
this can be extracted from the interference experiment. 
Our proposal for measuring the phase jump factor is to use the interference of BECs with internal degrees of freedom with Raman pulse and rf pulse, based on the study to observe the supercurrent decay in an annular BEC~\cite{Moulder2012}.

We consider atoms with two internal states $\ket{\uparrow}$
and $\ket{\downarrow}$.
First, prepare a BEC with the state $\ket{\uparrow}$, and create a soliton with the phase imprinting method~\cite{Denschlag2000}. 
Then, by applying the Raman $\pi/2$ pulse with a kick, half of the population of $\ket{\uparrow}$ is transferred into the state $\ket{\downarrow}$. 
After the kicked BEC with $\ket{\downarrow}$ moves with the distance $d$, then apply rf $\pi/2$ pulse, where half of populations in states $\ket{\uparrow}$ and $\ket{\downarrow}$ are converted into $\ket{\downarrow}$ and $\ket{ \uparrow}$, respectively. 
Then, measure the density profiles with the interference between BECs with and without displacement. 

The condensate wavefunctions after the Raman and rf pulses are given by 
\begin{align}
    \psi_\uparrow (x,y) = & \frac{1}{2} 
    [\psi_{\bf C} (x+d/2,y) - \psi_{\bf C} (x-d/2,y)], 
    \\
    \psi_\downarrow (x,y) = & \frac{1}{2} 
    [\psi_{\bf C} (x+d/2,y) + \psi_{\bf C} (x-d/2,y)], 
\end{align}
where we have shifted the $x$-axis with $d/2$ just for the consistency of the representation of the phase jump factor $S(x,y)$ in \eqref{eqS}. 
We then introduce the interference density ratio $R_\uparrow (x,y) \equiv  n_\uparrow (x,y)/[n_\uparrow (x,y)+n_\downarrow (x,y)]$. 
By using the relation $n_{i} (x,y) =  | \psi_{i} (x,y) |^2$ for $i = \uparrow, \downarrow$, we can reduce the interference density ratio as 
\begin{align}
    R_\uparrow (x,y) 
    = & 
    \frac{1}{2} 
    \left [ 1 - \chi (x,y) \cos (\Delta \theta ) \right]. 
    \label{eq17}
\end{align}
Here, the factor $\chi$ can be expressed by the ratio of the arithmetic and geometric means of density, given by 
\begin{align}
    \chi(x,y) \equiv & 
    \frac{2 
        | \psi_{\bf C} (x+d/2,y)| 
        | \psi_{\bf C} (x-d/2,y)| 
    }{
        | \psi_{\bf C} (x+d/2,y)|^2 
        +
        | \psi_{\bf C} (x-d/2,y)|^2 
    }. 
    \label{eq18}
\end{align}
Because of the inequality of arithmetic and geometric means, the mean ratio ranges as $0 \leq \chi (x,y) \leq 1$.

From Eq. \eqref{eq17}, 
the interference density ratio ranges as 
\begin{align}
   \frac{1-\chi(x,y)}{2}  \leq R_\uparrow (x,y) \leq \frac{1 + \chi(x,y)}{2} . 
\end{align}
In particular, if a soliton exists along the line with $(x_s, y_s)$, it is plausible that $|\psi_{\bf C} (x_s + d/2,y_s)| = | \psi_{\bf C} (x_s-d/2,y_s)|$ holds, if we take the distance $d$ as an appropriate value, such as a healing length $d = \xi$. 
This situation gives results such that $\chi (x_s, y_s) = 1$, $ 0 \leq R_\uparrow (x_s, y_s) \leq 1$, as well as $R_\uparrow (x_s, y_s ) = S (x_s, y_s) = [1 - \cos (\Delta \theta)]/2$, 
which implies that the interference density ratio $R_\uparrow$ possibly reproduces the phase jump factor $S$.  
The large value of the phase jump factor $S(x,y)$ and the interference density ratio $R_\uparrow (x,y)$ is due to the large phase difference $\Delta \theta$ and the large mean ratio $\chi(x,y)$, i.e., $|\psi_{\bf C} (x + d/2,y)| \simeq | \psi_{\bf C} (x-d/2,y)|$.

Interestingly, the mean ratio $\chi$ does not significantly contribute to the interference density ratio, and the profile of $R_\uparrow (x,y)$ can indeed reproduce the profile of the phase jump factor $S (x,y)$ very well even outside the region very close to the soliton (Fig.~\ref{p:hikabu}). 
Although the density profile of BECs itself is useful for detecting the valley of the density in the soliton, the interference density ratio $R_\uparrow$ is also useful because it allows us to see the phase jump characteristic of the soliton of BECs, and the breakdown of the soliton into the vortices.

\begin{figure*}[tbp]
\begin{center}
\includegraphics[clip,width=16.0cm]{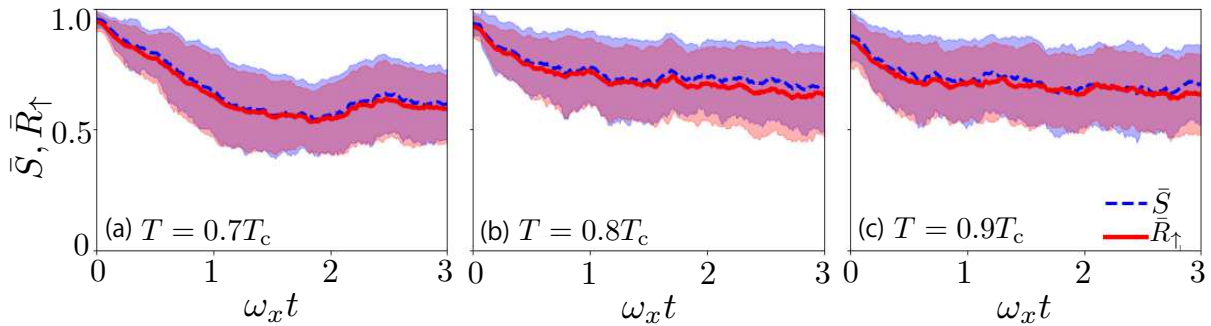}
\end{center}
\caption{
The time dependence of the soliton-path-averaged phase jump factor $\bar S$ (dashed line) and the soliton-path-averaged interference density ratio $\bar R_\uparrow$ (solid line) with the 100 sample average. The translucent regions show their variances. We used the same parameters as in Fig.~\ref{p:hikabu}. 
}
\label{p:hikati}
\end{figure*}

We can also confirm this fact through the mean values of the interference density ratio $R_\uparrow$ and the phase jump factor $S$ along the soliton path (Fig.~\ref{p:hikati}). 
The concept of the soliton path is defined as follows; 
If a dark soliton exists, the phase jump factor $S(x,y)$ can be maximum along the $x$-axis. 
If we accumulate the data of the position $( \mathop{\arg\min}\limits_{x \in X} [S (x,y)] ,y)$ within a certain region $X$ for $x$ (See Appendix~\ref{SecPath}), we can estimate the soliton path $l_{S}$ from the phase jump factor $S$, where the soliton path $l_S$ gives the structure of the dark soliton. 
Once the soliton path $l_{\rm S}$ is determined, 
the path-averaged phase jump factor $\bar S$ is obtained as 
\begin{align}
    \bar S = \frac{1}{L_{S}} \int_{l_{S}}  S(x,y) dl, 
\end{align}
where $L_{S}$ is the length of the path $l_{ S}$. 
So is the path-averaged interference density ratio, given by 
\begin{align}
    \bar R_\uparrow = 
    \frac{1}{L_{{R_\uparrow}}} 
    \int_{l_{R_\uparrow}} 
    R_\uparrow (x,y) d l, 
\end{align}
where $L_{{R_\uparrow}}$ is the path length of $l_{R_\uparrow}$.

\begin{figure}[tbp]
    \begin{center}
    \includegraphics[clip,width=8.0cm]{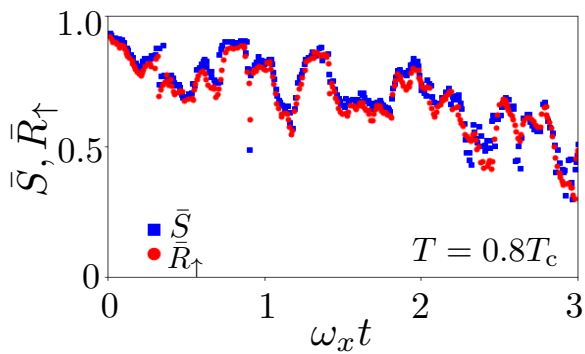}
    \end{center}
    \caption{
        The time dependence of the soliton-path-averaged phase jump factor $\bar S$ (square) and the soliton-path-averaged interference density ratio $\bar R_\uparrow$ (circle) in a single simulation. We used the same parameters as in Fig.~\ref{p:hikabu}. 
    }
    \label{fig4}
    \end{figure}

The distribution and the mean value of $R_\uparrow$ can reproduce the profile of $S$ very well (Figs.~\ref{p:hikabu} and \ref{p:hikati}). 
In Fig.~\ref{p:hikati}, the variances of $\bar S$ and $\bar R_\uparrow$ become large and saturate with the time evolution. 
This does not mean that the density interference ratio $R_\uparrow$ cannot sufficiently capture the phase jump factor $ S$. 
As in Fig.~\ref{fig4}, the time-dependence of $\bar R_\uparrow$ can also reproduce well that of $\bar S$ in a single run simulation. 
The large variance rather originates from the stochastically prepared initial state for the SPGPE. We start the simulation of the soliton dynamics from the equilibrium state, where the equilibrium is judged from the time-dependence of the number of particles in the coherent region $N_{\bf C}$ and the condensate $N_0$ in the SPGPE simulation as discussed Appendix~\ref{method}. 
This initial state has randomness in the classical field, and the time evolution of the phase jump factor $S$ and the interference density ratio $R_\uparrow$ strongly depends on the initial state. 
Therefore, the interference density ratio is essentially useful for studying the phase jump of the soliton and the breakdown of the soliton into vortices.

The present study focuses on the dark soliton with the phase difference $\pi$. It will be interesting to study the fate of the gray soliton. In the case of the gray soliton, the density (the amplitude of the condensate wave function) is symmetric along the center of the soliton~\cite{Tsuzuki:1971vw}, our method using the interference density ratio $R_\uparrow$ will work to reproduce the phase jump factor $S$ because of the relations~\eqref{eq17} and~\eqref{eq18}. 
However, searching the soliton path should be improved. In the dark soliton case, we have used the range given by the healing length $\xi$ for searching the soliton path as in Appendix~\ref{SecPath}. 
In the gray soliton case, the search range should become wider at least given by the soliton width $\xi_{\rm s} \equiv = \xi /\sqrt{1-(v_{\rm s}/c)^2}$~\cite{Tsuzuki:1971vw,Aycock:2017eu}, where $v_{\rm s}$ and $c$ is the speed of the soliton and Bogoliubov phonon, respectively.

As seen in Fig.~\ref{p:hikati}, $\bar R_\uparrow$ and $\bar S$ decay more slowly in the lower temperature case, where the order of the decay time is $1/\omega_x$. 
However, these values do not decay in the long-time regime and remain finite.
This long-time behavior is an artifact due to our method of tracing the soliton path in Appendix~\ref{SecPath}; this method always creates a line where the phase jump factor or the interference density ratio is maximum within a certain region, even if the soliton is no longer present.  
This difficulty could be removed by using machine learning techniques to trace the structure of solitons~\cite{FriederikeMetz:2021bh,ShangjieGuo:2021iy}. 
It will be interesting and useful to apply the machine learning technique to analyze the density and interference density ratio profile for understanding the soliton dynamics in BESs breakdown into vortices at nonzero temperatures.

\begin{figure*}[tbp]
\begin{center}
\includegraphics[clip,width=13.0cm]{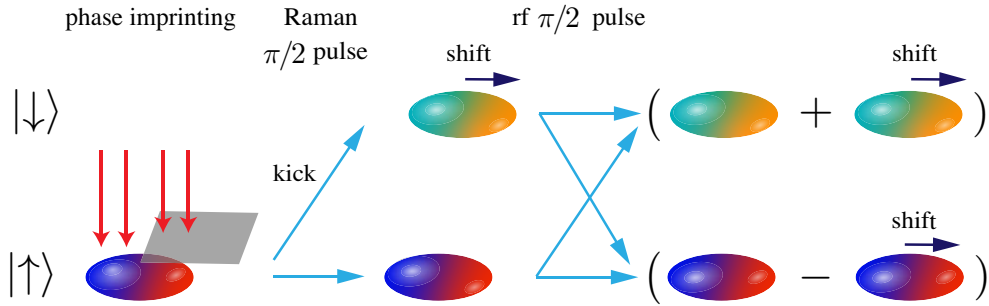}
\end{center}
\caption{
Method for measuring the interference density ratio to study the phase jump factor. 
We used two-component BECs with two internal states
$\ket{\uparrow}$ and $\ket{\downarrow}$. We first prepare the BEC and apply the phase imprinting to create soliton in the state $\ket{\uparrow}$. Then apply the Raman $\pi/2$ pulse to generate the kicked BEC in the state $\ket{\downarrow}$. 
After the kicked BEC has moved by the healing length, apply the rf $\pi/2$ pulse. 
Measuring the density profiles of BECs in both $\ket{\uparrow}$ and $\ket{\downarrow}$ states provides the interference density ratio $R_\uparrow$, which well reproduces the phase jump factor $S$. 
}
\label{p:kanji}
\end{figure*}

\section{conclusions}

We proposed a method to study the decay and breakdown of a phase imprinted soliton in an atomic Bose--Einstein condensate (BEC). 
First, we introduced the phase jump factor that measures the phase difference between two points. 
Using the (stochastic) projected Gross--Pitaevskii equation, we then demonstrated that the interference density ratio, which is the density ratio of two-component BECs after Raman and rf pulses are applied, can reproduce the phase jump factor very well. 
Our method will become an alternative method to study the soliton dynamics in BECs. 
The method used in this study to trace the soliton structure could be replaced with machine learning techniques.

\begin{acknowledgments}
The author has been supported by JSPS KAKENHI Grant No. JP16K17774. 
\end{acknowledgments} 


\appendix

\section{Method of simulation}\label{method} 

The stationary state of the BEC before applying the phase imprint method is prepared as follows.
Basically, the SPGPE provides the thermal equilibrium state at a given chemical potential and temperatures for an arbitrary initial state. 
However, to save the numerical simulation time and avoid accidentally reaching an unwanted state, 
we start the simulation with a state obtained from the mean-field approximation, which is close to the thermal equilibrium state for the SPGPE. 

In the mean-field approximation, 
we first suppose that the number density in the coherent region can be given by the Thomas-Fermi (TF) approximation, given by 
$n_{{\bf C}}^{\rm TF} (x,y) =   [ \mu - m (\omega_x^2 x^2 + \omega_y^2 y^2)/2 ]/g$ for $ \mu \geq m (\omega_x^2 x^2 + \omega_y^2 y^2)/2$ and $n_{{\bf C}}^{\rm TF} (x,y) = 0$ for otherwise. 
The number of particles in the coherent region ${\bf C}$ and in the incoherent region ${\bf IC}$ in the TF approximation are then given by 
\begin{align}
    N_{{\bf C}}^{{\rm TF}}  = & \frac{\pi\mu^2}{m g \omega_x\omega_y}, 
    \\ 
    N_{{\bf IC}}^{{\rm TF}} = & \int_{\epsilon_{\rm cut}}^\infty 
    d \epsilon \frac{\rho_{\rm HF} (\epsilon)}{e^{\beta (\epsilon - \mu)} - 1}.  
\end{align}
Here, $\rho_{\rm HF}$ is the density of states in the semi-classical Hartree--Fock approximation given by 
\begin{align}
    \rho_{\rm HF} (\epsilon)
    = & 
    \int \frac{d {\bf r}d {\bf p}}{(2\pi \hbar)^2} 
    \delta 
    \biggl ( 
        \epsilon - \biggl [ 
            \frac{{\bf p}^2}{2m} 
            \nonumber \\ & 
            + \frac{m}{2} (\omega_x^2 x^2 + \omega_y^2 y^2) + 2 g n_{{\bf C}}^{{\rm TF}} (x,y)
             \biggr ]
    \biggr ), 
\end{align}
where we assumed that the particle density in the incoherent region is much lower than that in the coherent region: $n_{\bf IC} \ll n_{\bf C}$. 
The density of states in the two-dimensional harmonic trap system is given by 
\begin{align}
    \rho_{\rm HF} (\epsilon) = 
    \begin{cases}
        \cfrac{2 (\epsilon-\mu)}{\hbar^2 \omega_x \omega_y}, & \qquad \mu \leq \epsilon < 2 \mu , 
        \\
        \cfrac{\epsilon}{\hbar^2 \omega_x \omega_y}. 
        & \qquad 2 \mu \leq \epsilon. 
    \end{cases}
\end{align}

The energy cutoff $\epsilon_{\rm cut}$ for separating the coherent and incoherent regions is determined by using the occupation number with the Bose--Einstein distribution function for an ideal Bose gas, where the cutoff occupation number $N_{\rm cut}$ is given by $N_{\rm cut} = 1/[e^{\beta (\epsilon_{\rm cut} - \mu)}-1]$. 
The energy cutoff is then given in the form 
\begin{align}
    \epsilon_{\rm cut} = \mu + k_{\rm B} T \ln \left ( 1 + \frac{1}{N_{\rm cut}} \right ). 
\end{align}
In this study, the cutoff occupation number is set as $N_{\rm cut} = 3$, where the Planck law is well approximated by the Rayleigh--Jeans law for the black body radiation~\cite{Blakie2007}. 
Given the temperature and chemical potential, then the energy cutoff is determined.

\begin{figure}[tbp]
\begin{center}
\includegraphics[clip,width=8.0cm]{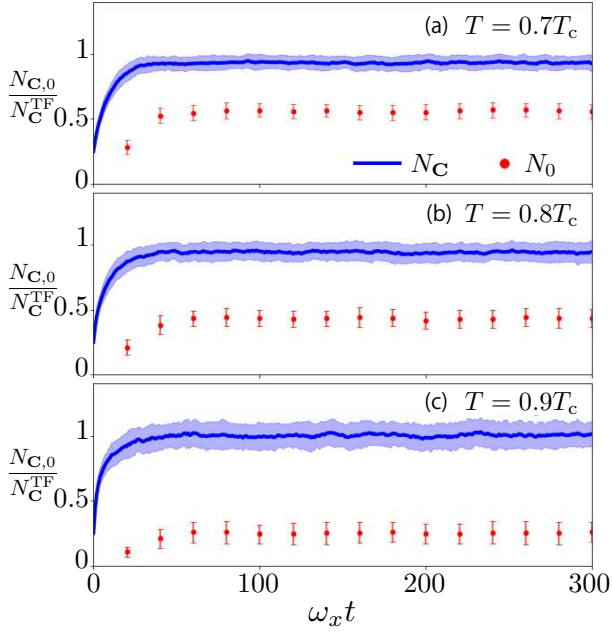}
\end{center}
\caption{
Time evolution of the number of particles in the coherent region $N_{\bf C}$ and the condensate $N_0$ simulated by SPGPE, scaled by the number of particles in the coherent region $N_{\bf C}^{\rm TF}$ 
in the Thomas--Fermi and Hartree--Fock approximations.
We used $\omega_y/\omega_x=5$ and 100 sample averages. 
For evaluating the condensate number of particles $N_0$, we take the time average with a period of $20\omega_x t$. 
At all temperatures in the simulation, the behavior of $N_{{\bf C},0}$ indicates that the system reaches equilibrium at $\omega_x t \sim 100$.
In order to highlight the thermal equilibration process in SPGPE, we replace $c_{\nu} (t=0)$ in Eq.~\eqref{c_nu_initial} with $c_{\nu} (t=0)/4$, where the initial condition of the number of the particles in the coherent region is set to $N_{{\bf C}}^{\rm TF}/4$. 
}
\label{p:Nhattenh}
\end{figure}

As a preprocessing of the simulation for the SPGPE, we first set the number of particles $N_{\rm tot}$ and the temperature $T$. Then the chemical potential $\mu$ is determined from the relation $N_{\rm tot} = N_{{\bf C}}^{{\rm TF}} + N_{{\bf IC}}^{{\rm TF}}$ within the Thomas--Fermi approximation and semi-classical Hartree--Fock approximation, which provides the number of the particles in the coherent region $N_{\bf C}$ as well as $\epsilon_{\rm cut}$. 
We can also evaluate the healing length $\xi = \hbar /\sqrt{2m\mu}$ and the sound velocity $v = \hbar /(\sqrt{2} m \xi)$. 
Once the number of the particles in the coherent region is determined, we consider the relation 
\begin{align}
    \sum\limits_{\nu \in {\bf C}} \frac{1}{e^{\beta (\epsilon_{\nu} - \mu')} - 1} = N_{{\bf C}}^{{\rm TF}}, 
\end{align}
where $\epsilon_{\nu} = \hbar \omega_x \left ( n_x + 1/2 \right ) + \hbar \omega_y \left ( n_y + 1/2 \right )$ with $\nu = (n_x,n_y)$ and $\mu'$ is the artificial chemical potential introduced to fix the number of the particles in the coherent region to be $N_{\bf C}^{\rm TF}$. 
After determining $\mu'$ to satisfy the above condition, 
we take the initial condition of the coefficient in Eq. \eqref{titen} as 
\begin{align}
    c_{\nu} (t=0) = \frac{1}{\sqrt{e^{\beta (\epsilon_{\nu} - \mu')} - 1}} e^{\theta_{\nu}}, 
    \label{c_nu_initial}
\end{align}
where $\theta_{\nu}$ is the phase randomly generated in the range $ [-\pi,\pi)$.

In order to check the stationary state, 
we monitor the time-dependence of the number of particles in the coherent region $N_{{\bf C}}$, 
and the condensate number of particles $N_{0}$. 
According to the criterion provided by Penrose and Onsager~\cite{Yang1962,Penrose1956}, the condensate number of particles is given by the maximum eigenvalue of the one-body density matrix
\begin{align}
    \rho_1 ({\bf r}_1, {\bf r}_2) \equiv & 
    \langle 
    \hat \psi_{\bf C}^\dag ({\bf r}_1)
    \hat \psi_{\bf C} ({\bf r}_2)
    \rangle 
    \\ 
    = & 
    \sum\limits_{\nu,\nu' \in {\bf C}} n_{\nu, \nu'} 
    u_{\nu}^* ({\bf r}_1) u_{\nu'} ({\bf r}_2)
\end{align}
where we have used $\hat \psi_{\bf C} ({\bf r}) = \sum\limits_{\nu \in {\bf C}} u_\nu ({\bf r}) \hat a_\nu$ and $n_{\nu,\nu'} \equiv \langle \hat a_\nu^\dag \hat a_\nu '\rangle$. 
To consider the stationary state in the single run simulation, we assume the Ergodic hypothesis where the statistical average is replaced with the long time average, i.e., $n_{\nu,\nu'} = \overline{c_\nu^* c_{\nu '}}$. 
We then diagonalize the matrix $n_{\nu,\nu'}$, whose maximum eigenvalue gives the condensate number of particles $N_0$. (See Fig.~\ref{p:Nhattenh}.)

\section{Analytic relation in the phase imprinting}\label{AppendixB}

We label $t=0^-$ and $t=0^+$ as the time just before and after applying the phase imprinting method to the steady-state BEC.
The field is given by 
\begin{align}
    \tilde \psi_{\bf C}^\pm (X,Y) \equiv & \tilde \psi_{\bf C} (X,Y, \tau = 0^\pm) 
    \\ 
    = & \sum\limits_{n,l} c_{n,l}^{\pm} \tilde \phi_n (X) \tilde \phi_l (Y), 
\end{align}
which are related as 
\begin{align}
    \tilde \psi_{\bf C}^+ (X,Y) = 
    \begin{cases}
        \tilde \psi_{\bf C}^- (X,Y), & X > 0,
        \\
        e^{i \theta_{\rm i}} \tilde \psi_{\bf C}^- (X,Y) & X < 0.  
    \end{cases}
\end{align}
Here we assumed the imprinted phase jump is $\theta_{\rm i}$ and used the normalized notation such as $\tau \equiv \omega_x t$, $\tilde \psi_{\bf C} \equiv \psi_{\bf C}/\sqrt{N_{\bf C} /a_{{\rm ho},x}^2}$, $X \equiv x/a_{{\rm ho},x}$ and $Y \equiv y /a_{{\rm ho},x}$ with $a_{{\rm ho},x} \equiv \sqrt{\hbar / (m\omega_x)}$. 

The coefficient $c_{n,l}^+$ after the phase imprinting is related to the coefficient $c_{n,l}^-$ just before the phase imprinting through
\begin{align}
    c_{n,l}^{+} = \sum\limits_{n' \in {\bf C}} c_{n',l}^- 
    [ I_{nn'}^{(+)} + e^{i \theta_{\rm i} } I_{nn'}^{(-)}], 
\end{align}
where 
\begin{align}
    I_{n,n'}^{(\pm)} \equiv \pm \int_0^{\pm \infty} d X \tilde \phi_n^* (X) \tilde \phi_{n'} (X). 
\end{align}
Using the properties of the Hermite polynomials, we can give the analytic expressions of $I_{n,n'}^{(\pm)}$ as 
\onecolumngrid
\begin{align}
    I_{n,n'}^{(\pm)} = 
    \begin{cases}
        1/2, & n = n',
        \\
        \pm \cfrac{1}{\sqrt{\pi 2^{n+n'} n! n'!}} \cfrac{H_n (0) H_{n'+1}(0)}{2 (n-n')}, & 
         (n,n') = ({\rm even},{\rm odd}),
                       \\
        \pm \cfrac{1}{\sqrt{\pi 2^{n+n'} n! n'!}} \cfrac{H_{n+1} (0) H_{n'}(0)}{2 (n'-n)}, & (n,n') = ({\rm odd},{\rm even}),
        \\
        0, & {\rm otherwise}.
    \end{cases}
\end{align}

\twocolumngrid

\section{Method to determine the soliton path}\label{SecPath}

This Appendix explains the method to determine the soliton path used in this paper.
A basic strategy is to connect the points with large values of the phase jump factor or the interference density along the $y$-axis. 
In detail, the method is slightly different when the soliton is created and after starting the time-evolution of the soliton dynamics. 

\begin{figure*}[tbp]
\begin{center}
\includegraphics[clip,width=15.0cm]{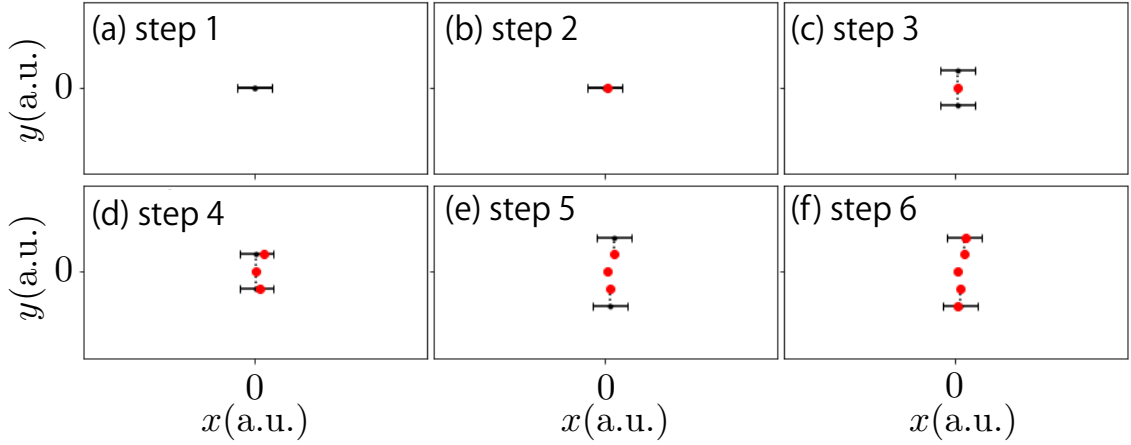}
\end{center}
\caption{Method to determine the soliton path at $t = 0$. 
(a) Search for the maximum value of $S$ or $R_\uparrow$ with the range $X = (- \xi, + \xi)$ at $y=0$. The bar shows the search range of the maximum value. 
(b) Mark the maximum value position (red point). Suppose this point is $(x_{\rm max}(y=0), y=0)$ .
(c) Search for the maximum value of $S$ or $R_\uparrow$ at $y=\pm \Delta y$ with the range $X = (x_{\rm max}(0) - \xi, x_{\rm max} (0) + \xi)$, which is represented by bars. 
(d) Mark the maximum value position (red point). Suppose this point is $x_{\rm max}(\pm \Delta y)$ 
(e) Search for the maximum value of $S$ or $R_\uparrow$ at $y=\pm 2 \Delta y$ in the range $X = (x_{\rm max}(\pm \Delta y) - \xi, x_{\rm max} (\pm \Delta y) + \xi)$. 
(d) Mark the maximum value position (red point). Suppose this point is $x_{\rm max}(\pm 2 \Delta y)$ 
}
\label{p:mapa0}
\end{figure*}

Just after the phase jump of $\pi$ is imprinted at $x=0$, i.e., at $t=0^+$, 
we search for the position where $S$ or $R_\uparrow$ has the maximum value near $x=0$ in the region $X$, which will be explained below. 
In the numerical simulation, we actually take the mesh in the coordinate space (such as with a grid spacing $\Delta y$ in the $y$-direction). 
If the maximum value of $S$ or $R_\uparrow$ is found at $(x_{\rm max} (y), y)$, 
then we search for the soliton position at $y+\Delta y$ in the range $X = (x_{\rm max} (y) - \xi,  x_{\rm max}(y) + \xi )$. 
We start the search of the maximum value of $S$ or $R_\uparrow$ at $x = y = 0$. 
(See Fig.~\ref{p:mapa0}). 

For the time-evolving soliton, the soliton position at the time $t + \Delta t$ is searched based on the preceding soliton path data at the time $t$. 
If the soliton path is given by the function $x(t;y)$, 
we search for the maximum value of $S$ or $R_\uparrow$ within the region $ X = (x(t;y) - v \Delta t, x(t;y) + v \Delta t)$ to find the soliton path at $t+\Delta t$, where $v$ is the sound velocity, which gives the upper limit of the speed of the soliton at $T=0$. 
In this method, it sometimes happens that the path is disconnected, as shown in Fig.~\ref{p:shpa}. 
In this case, we reconstruct the path starting from the disconnected point by using the healing length, such as $X = (x(t;y) - \xi, x(t;y) + \xi)$ for the soliton path at $t+\Delta t$ (See Fig.~\ref{p:shpa}).

\begin{figure*}[tbp]
\begin{center}
\includegraphics[clip,width=11.0cm]{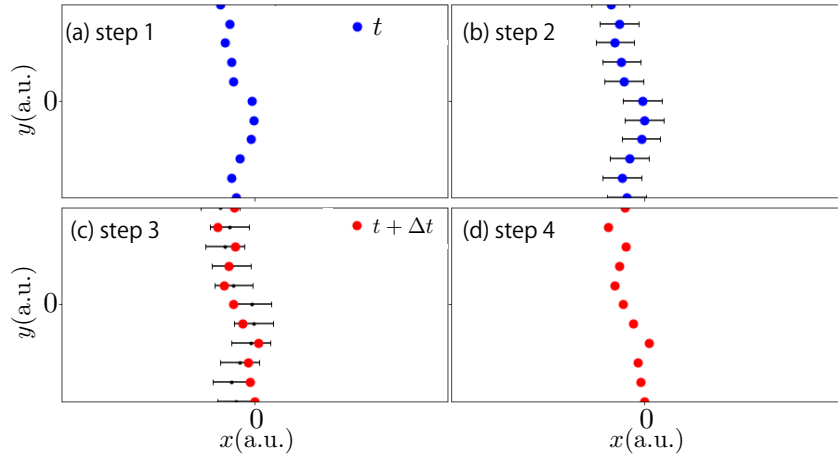}
\end{center}
\caption{
Method to determine the soliton path at the time $t + \Delta t$ based on the data at the time $t$. 
(a) Prepare the soliton path at $t$, which is constructed by the data $(x_{\rm max} (t;y), y)$ (blue points). 
(b) Search for the maximum value of $S$ or $R_\uparrow$ in the range $X = (x_{\rm max} (t;y) - v \Delta, x_{\rm max} (t;y) + v \Delta t)$ at $y$, where $v$ is the sound velocity, which gives the upper limit of the speed of the soliton at $T=0$. Bars represent the region $X$ to search for the maximum value. 
(c) Mark the maximum value position, which gives $(x_{\rm max}(t+\Delta t; y), y)$ (red points). 
(d) This data set gives the soliton path at $t + \Delta t$. 
}
\label{p:mapa1}
\end{figure*}

\begin{figure}[tbp]
\begin{center}
\includegraphics[clip,width=9.0cm]{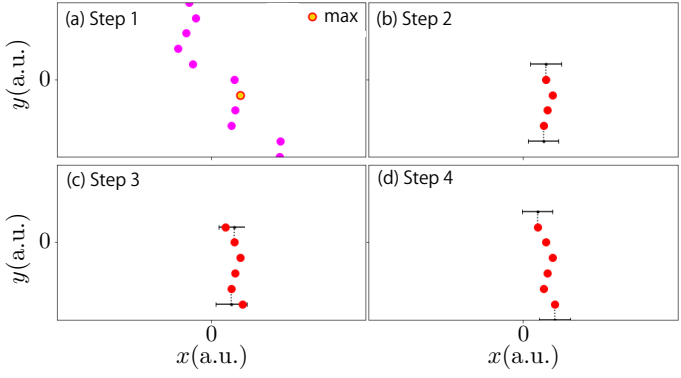}
\end{center}
\caption{
Method to correct the disconnected soliton path. Disconnection is determined by the criteria $| x_{\rm max} (y+\Delta y) - x_{\rm max} (y) | > \xi$. 
(a) Find the maximum value of $S$ or $R_\uparrow$ in the soliton path $(x_{\rm max} (t;y), y)$, which becomes a start point $(x_{\rm max} (t;y_0) , y_0)$ for correcting the soliton path (a highlighted point). 
(b) Search for the end points of a part of the soliton line starting from $(x_{\rm max} (t;y_0) , y_0)$, where, for example, the upper end position is supposed to be $(x_{\rm max} (y'), y')$ that satisfies $| x_{\rm max} (y'+\Delta y) - x_{\rm max} (y') | > \xi$. 
(c) Search for the maximum value of $S$ or $R_\uparrow$ in the range $X = (x_{\rm max} (y') - \xi, x_{\rm max} (y') + \xi)$ at $y' + \Delta y$, and mark the position (red points). 
(d) Continue the same search process as shown in Fig.~\ref{p:mapa0} for $t=0$. 
}
\label{p:shpa}
\end{figure}

\bibliographystyle{apsrev4-2}
\bibliography{draft.bib}

\begin{thebibliography}{34}%
\makeatletter
\providecommand \@ifxundefined [1]{%
 \@ifx{#1\undefined}
}%
\providecommand \@ifnum [1]{%
 \ifnum #1\expandafter \@firstoftwo
 \else \expandafter \@secondoftwo
 \fi
}%
\providecommand \@ifx [1]{%
 \ifx #1\expandafter \@firstoftwo
 \else \expandafter \@secondoftwo
 \fi
}%
\providecommand \natexlab [1]{#1}%
\providecommand \enquote  [1]{``#1''}%
\providecommand \bibnamefont  [1]{#1}%
\providecommand \bibfnamefont [1]{#1}%
\providecommand \citenamefont [1]{#1}%
\providecommand \href@noop [0]{\@secondoftwo}%
\providecommand \href [0]{\begingroup \@sanitize@url \@href}%
\providecommand \@href[1]{\@@startlink{#1}\@@href}%
\providecommand \@@href[1]{\endgroup#1\@@endlink}%
\providecommand \@sanitize@url [0]{\catcode `\\12\catcode `\$12\catcode
  `\&12\catcode `\#12\catcode `\^12\catcode `\_12\catcode `\%12\relax}%
\providecommand \@@startlink[1]{}%
\providecommand \@@endlink[0]{}%
\providecommand \url  [0]{\begingroup\@sanitize@url \@url }%
\providecommand \@url [1]{\endgroup\@href {#1}{\urlprefix }}%
\providecommand \urlprefix  [0]{URL }%
\providecommand \Eprint [0]{\href }%
\providecommand \doibase [0]{https://doi.org/}%
\providecommand \selectlanguage [0]{\@gobble}%
\providecommand \bibinfo  [0]{\@secondoftwo}%
\providecommand \bibfield  [0]{\@secondoftwo}%
\providecommand \translation [1]{[#1]}%
\providecommand \BibitemOpen [0]{}%
\providecommand \bibitemStop [0]{}%
\providecommand \bibitemNoStop [0]{.\EOS\space}%
\providecommand \EOS [0]{\spacefactor3000\relax}%
\providecommand \BibitemShut  [1]{\csname bibitem#1\endcsname}%
\let\auto@bib@innerbib\@empty
\bibitem [{\citenamefont {Su}\ \emph {et~al.}(1979)\citenamefont {Su},
  \citenamefont {Schrieffer},\ and\ \citenamefont {Heeger}}]{Su1698}%
  \BibitemOpen
  \bibfield  {author} {\bibinfo {author} {\bibfnamefont {W.~P.}\ \bibnamefont
  {Su}}, \bibinfo {author} {\bibfnamefont {J.~R.}\ \bibnamefont {Schrieffer}},\
  and\ \bibinfo {author} {\bibfnamefont {A.~J.}\ \bibnamefont {Heeger}},\
  }\href@noop {} {\bibfield  {journal} {\bibinfo  {journal} {Phys. Rev. Lett.}\
  }\textbf {\bibinfo {volume} {42}},\ \bibinfo {pages} {1698} (\bibinfo {year}
  {1979})}\BibitemShut {NoStop}%
\bibitem [{\citenamefont {Ikezi}\ \emph {et~al.}(1970)\citenamefont {Ikezi},
  \citenamefont {Taylor},\ and\ \citenamefont {Baker}}]{Ikezi1970}%
  \BibitemOpen
  \bibfield  {author} {\bibinfo {author} {\bibfnamefont {H.}~\bibnamefont
  {Ikezi}}, \bibinfo {author} {\bibfnamefont {R.~J.}\ \bibnamefont {Taylor}},\
  and\ \bibinfo {author} {\bibfnamefont {D.~R.}\ \bibnamefont {Baker}},\
  }\href@noop {} {\bibfield  {journal} {\bibinfo  {journal} {Phys. Rev. Lett.}\
  }\textbf {\bibinfo {volume} {25}},\ \bibinfo {pages} {11} (\bibinfo {year}
  {1970})}\BibitemShut {NoStop}%
\bibitem [{\citenamefont {Maxworthy}\ and\ \citenamefont
  {Redekopp}(1976)}]{MAXWORTHY1976261}%
  \BibitemOpen
  \bibfield  {author} {\bibinfo {author} {\bibfnamefont {T.}~\bibnamefont
  {Maxworthy}}\ and\ \bibinfo {author} {\bibfnamefont {L.}~\bibnamefont
  {Redekopp}},\ }\href@noop {} {\bibfield  {journal} {\bibinfo  {journal}
  {Icarus}\ }\textbf {\bibinfo {volume} {29}},\ \bibinfo {pages} {261}
  (\bibinfo {year} {1976})}\BibitemShut {NoStop}%
\bibitem [{\citenamefont {Burger}\ \emph {et~al.}(1999)\citenamefont {Burger},
  \citenamefont {Bongs}, \citenamefont {Dettmer}, \citenamefont {Ertmer},
  \citenamefont {Sengstock}, \citenamefont {Sanpera}, \citenamefont
  {Shlyapnikov},\ and\ \citenamefont {Lewenstein}}]{Burger1999}%
  \BibitemOpen
  \bibfield  {author} {\bibinfo {author} {\bibfnamefont {S.}~\bibnamefont
  {Burger}}, \bibinfo {author} {\bibfnamefont {K.}~\bibnamefont {Bongs}},
  \bibinfo {author} {\bibfnamefont {S.}~\bibnamefont {Dettmer}}, \bibinfo
  {author} {\bibfnamefont {W.}~\bibnamefont {Ertmer}}, \bibinfo {author}
  {\bibfnamefont {K.}~\bibnamefont {Sengstock}}, \bibinfo {author}
  {\bibfnamefont {A.}~\bibnamefont {Sanpera}}, \bibinfo {author} {\bibfnamefont
  {G.~V.}\ \bibnamefont {Shlyapnikov}},\ and\ \bibinfo {author} {\bibfnamefont
  {M.}~\bibnamefont {Lewenstein}},\ }\href
  {https://doi.org/10.1103/PhysRevLett.83.5198} {\bibfield  {journal} {\bibinfo
   {journal} {Phys. Rev. Lett.}\ }\textbf {\bibinfo {volume} {83}},\ \bibinfo
  {pages} {5198} (\bibinfo {year} {1999})}\BibitemShut {NoStop}%
\bibitem [{\citenamefont {Denschlag}\ \emph {et~al.}(2000)\citenamefont
  {Denschlag}, \citenamefont {Simsarian}, \citenamefont {Feder}, \citenamefont
  {Clark}, \citenamefont {Collins}, \citenamefont {Cubizolles}, \citenamefont
  {Deng}, \citenamefont {Hagley}, \citenamefont {Helmerson}, \citenamefont
  {Reinhardt}, \citenamefont {Rolston}, \citenamefont {Schneider},\ and\
  \citenamefont {Phillips}}]{Denschlag2000}%
  \BibitemOpen
  \bibfield  {author} {\bibinfo {author} {\bibfnamefont {J.}~\bibnamefont
  {Denschlag}}, \bibinfo {author} {\bibfnamefont {J.~E.}\ \bibnamefont
  {Simsarian}}, \bibinfo {author} {\bibfnamefont {D.~L.}\ \bibnamefont
  {Feder}}, \bibinfo {author} {\bibfnamefont {C.~W.}\ \bibnamefont {Clark}},
  \bibinfo {author} {\bibfnamefont {L.~A.}\ \bibnamefont {Collins}}, \bibinfo
  {author} {\bibfnamefont {J.}~\bibnamefont {Cubizolles}}, \bibinfo {author}
  {\bibfnamefont {L.}~\bibnamefont {Deng}}, \bibinfo {author} {\bibfnamefont
  {E.~W.}\ \bibnamefont {Hagley}}, \bibinfo {author} {\bibfnamefont
  {K.}~\bibnamefont {Helmerson}}, \bibinfo {author} {\bibfnamefont {W.~P.}\
  \bibnamefont {Reinhardt}}, \bibinfo {author} {\bibfnamefont {S.~L.}\
  \bibnamefont {Rolston}}, \bibinfo {author} {\bibfnamefont {B.~I.}\
  \bibnamefont {Schneider}},\ and\ \bibinfo {author} {\bibfnamefont {W.~D.}\
  \bibnamefont {Phillips}},\ }\href@noop {} {\bibfield  {journal} {\bibinfo
  {journal} {Science}\ }\textbf {\bibinfo {volume} {287}},\ \bibinfo {pages}
  {97} (\bibinfo {year} {2000})}\BibitemShut {NoStop}%
\bibitem [{\citenamefont {Dutton}\ \emph {et~al.}(2001)\citenamefont {Dutton},
  \citenamefont {Budde}, \citenamefont {Slowe},\ and\ \citenamefont
  {Hau}}]{Dutton2001}%
  \BibitemOpen
  \bibfield  {author} {\bibinfo {author} {\bibfnamefont {Z.}~\bibnamefont
  {Dutton}}, \bibinfo {author} {\bibfnamefont {M.}~\bibnamefont {Budde}},
  \bibinfo {author} {\bibfnamefont {C.}~\bibnamefont {Slowe}},\ and\ \bibinfo
  {author} {\bibfnamefont {L.~V.}\ \bibnamefont {Hau}},\ }\href@noop {}
  {\bibfield  {journal} {\bibinfo  {journal} {Science}\ }\textbf {\bibinfo
  {volume} {293}},\ \bibinfo {pages} {663} (\bibinfo {year}
  {2001})}\BibitemShut {NoStop}%
\bibitem [{\citenamefont {Brand}\ and\ \citenamefont
  {Reinhardt}(2002)}]{Brand2002}%
  \BibitemOpen
  \bibfield  {author} {\bibinfo {author} {\bibfnamefont {J.}~\bibnamefont
  {Brand}}\ and\ \bibinfo {author} {\bibfnamefont {W.~P.}\ \bibnamefont
  {Reinhardt}},\ }\href {https://doi.org/10.1103/PhysRevA.65.043612} {\bibfield
   {journal} {\bibinfo  {journal} {Phys. Rev. A}\ }\textbf {\bibinfo {volume}
  {65}},\ \bibinfo {pages} {043612} (\bibinfo {year} {2002})}\BibitemShut
  {NoStop}%
\bibitem [{\citenamefont {Anderson}\ \emph {et~al.}(2001)\citenamefont
  {Anderson}, \citenamefont {Haljan}, \citenamefont {Regal}, \citenamefont
  {Feder}, \citenamefont {Collins}, \citenamefont {Clark},\ and\ \citenamefont
  {Cornell}}]{Anderson2001}%
  \BibitemOpen
  \bibfield  {author} {\bibinfo {author} {\bibfnamefont {B.~P.}\ \bibnamefont
  {Anderson}}, \bibinfo {author} {\bibfnamefont {P.~C.}\ \bibnamefont
  {Haljan}}, \bibinfo {author} {\bibfnamefont {C.~A.}\ \bibnamefont {Regal}},
  \bibinfo {author} {\bibfnamefont {D.~L.}\ \bibnamefont {Feder}}, \bibinfo
  {author} {\bibfnamefont {L.~A.}\ \bibnamefont {Collins}}, \bibinfo {author}
  {\bibfnamefont {C.~W.}\ \bibnamefont {Clark}},\ and\ \bibinfo {author}
  {\bibfnamefont {E.~A.}\ \bibnamefont {Cornell}},\ }\href@noop {} {\bibfield
  {journal} {\bibinfo  {journal} {Phys. Rev. Lett.}\ }\textbf {\bibinfo
  {volume} {86}},\ \bibinfo {pages} {2926} (\bibinfo {year}
  {2001})}\BibitemShut {NoStop}%
\bibitem [{\citenamefont {Muryshev}\ \emph {et~al.}(2002)\citenamefont
  {Muryshev}, \citenamefont {Shlyapnikov}, \citenamefont {Ertmer},
  \citenamefont {Sengstock},\ and\ \citenamefont
  {Lewenstein}}]{PhysRevLett.89.110401}%
  \BibitemOpen
  \bibfield  {author} {\bibinfo {author} {\bibfnamefont {A.}~\bibnamefont
  {Muryshev}}, \bibinfo {author} {\bibfnamefont {G.~V.}\ \bibnamefont
  {Shlyapnikov}}, \bibinfo {author} {\bibfnamefont {W.}~\bibnamefont {Ertmer}},
  \bibinfo {author} {\bibfnamefont {K.}~\bibnamefont {Sengstock}},\ and\
  \bibinfo {author} {\bibfnamefont {M.}~\bibnamefont {Lewenstein}},\ }\href
  {https://doi.org/10.1103/PhysRevLett.89.110401} {\bibfield  {journal}
  {\bibinfo  {journal} {Phys. Rev. Lett.}\ }\textbf {\bibinfo {volume} {89}},\
  \bibinfo {pages} {110401} (\bibinfo {year} {2002})}\BibitemShut {NoStop}%
\bibitem [{\citenamefont {Ku}\ \emph {et~al.}(2014)\citenamefont {Ku},
  \citenamefont {Ji}, \citenamefont {Mukherjee}, \citenamefont
  {Guardado-Sanchez}, \citenamefont {Cheuk}, \citenamefont {Yefsah},\ and\
  \citenamefont {Zwierlein}}]{PhysRevLett.113.065301}%
  \BibitemOpen
  \bibfield  {author} {\bibinfo {author} {\bibfnamefont {M.~J.~H.}\
  \bibnamefont {Ku}}, \bibinfo {author} {\bibfnamefont {W.}~\bibnamefont {Ji}},
  \bibinfo {author} {\bibfnamefont {B.}~\bibnamefont {Mukherjee}}, \bibinfo
  {author} {\bibfnamefont {E.}~\bibnamefont {Guardado-Sanchez}}, \bibinfo
  {author} {\bibfnamefont {L.~W.}\ \bibnamefont {Cheuk}}, \bibinfo {author}
  {\bibfnamefont {T.}~\bibnamefont {Yefsah}},\ and\ \bibinfo {author}
  {\bibfnamefont {M.~W.}\ \bibnamefont {Zwierlein}},\ }\href
  {https://doi.org/10.1103/PhysRevLett.113.065301} {\bibfield  {journal}
  {\bibinfo  {journal} {Phys. Rev. Lett.}\ }\textbf {\bibinfo {volume} {113}},\
  \bibinfo {pages} {065301} (\bibinfo {year} {2014})}\BibitemShut {NoStop}%
\bibitem [{\citenamefont {Donadello}\ \emph {et~al.}(2014)\citenamefont
  {Donadello}, \citenamefont {Serafini}, \citenamefont {Tylutki}, \citenamefont
  {Pitaevskii}, \citenamefont {Dalfovo}, \citenamefont {Lamporesi},\ and\
  \citenamefont {Ferrari}}]{PhysRevLett.113.065302}%
  \BibitemOpen
  \bibfield  {author} {\bibinfo {author} {\bibfnamefont {S.}~\bibnamefont
  {Donadello}}, \bibinfo {author} {\bibfnamefont {S.}~\bibnamefont {Serafini}},
  \bibinfo {author} {\bibfnamefont {M.}~\bibnamefont {Tylutki}}, \bibinfo
  {author} {\bibfnamefont {L.~P.}\ \bibnamefont {Pitaevskii}}, \bibinfo
  {author} {\bibfnamefont {F.}~\bibnamefont {Dalfovo}}, \bibinfo {author}
  {\bibfnamefont {G.}~\bibnamefont {Lamporesi}},\ and\ \bibinfo {author}
  {\bibfnamefont {G.}~\bibnamefont {Ferrari}},\ }\href
  {https://doi.org/10.1103/PhysRevLett.113.065302} {\bibfield  {journal}
  {\bibinfo  {journal} {Phys. Rev. Lett.}\ }\textbf {\bibinfo {volume} {113}},\
  \bibinfo {pages} {065302} (\bibinfo {year} {2014})}\BibitemShut {NoStop}%
\bibitem [{\citenamefont {Cockburn}\ \emph {et~al.}(2011)\citenamefont
  {Cockburn}, \citenamefont {Nistazakis}, \citenamefont {P.Horikis},
  \citenamefont {G.Kevrekidis}, \citenamefont {P.Proukakis},\ and\
  \citenamefont {J.Frantzeskakis}}]{PhysRevA.84.043640}%
  \BibitemOpen
  \bibfield  {author} {\bibinfo {author} {\bibfnamefont {S.~P.}\ \bibnamefont
  {Cockburn}}, \bibinfo {author} {\bibfnamefont {H.~E.}\ \bibnamefont
  {Nistazakis}}, \bibinfo {author} {\bibfnamefont {T.}~\bibnamefont
  {P.Horikis}}, \bibinfo {author} {\bibfnamefont {P.}~\bibnamefont
  {G.Kevrekidis}}, \bibinfo {author} {\bibfnamefont {N.}~\bibnamefont
  {P.Proukakis}},\ and\ \bibinfo {author} {\bibfnamefont {D.}~\bibnamefont
  {J.Frantzeskakis}},\ }\href {https://doi.org/10.1103/PhysRevA.84.043640}
  {\bibfield  {journal} {\bibinfo  {journal} {Phys. Rev. A}\ }\textbf {\bibinfo
  {volume} {84}},\ \bibinfo {pages} {043640} (\bibinfo {year}
  {2011})}\BibitemShut {NoStop}%
\bibitem [{\citenamefont {Aycock}\ \emph {et~al.}(2017)\citenamefont {Aycock},
  \citenamefont {Hurst}, \citenamefont {Efimkin}, \citenamefont {Genkina},
  \citenamefont {Lu}, \citenamefont {Galitski},\ and\ \citenamefont
  {Spielman}}]{Aycock:2017eu}%
  \BibitemOpen
  \bibfield  {author} {\bibinfo {author} {\bibfnamefont {L.~M.}\ \bibnamefont
  {Aycock}}, \bibinfo {author} {\bibfnamefont {H.~M.}\ \bibnamefont {Hurst}},
  \bibinfo {author} {\bibfnamefont {D.~K.}\ \bibnamefont {Efimkin}}, \bibinfo
  {author} {\bibfnamefont {D.}~\bibnamefont {Genkina}}, \bibinfo {author}
  {\bibfnamefont {H.-I.}\ \bibnamefont {Lu}}, \bibinfo {author} {\bibfnamefont
  {V.~M.}\ \bibnamefont {Galitski}},\ and\ \bibinfo {author} {\bibfnamefont
  {I.~B.}\ \bibnamefont {Spielman}},\ }\href@noop {} {\bibfield  {journal}
  {\bibinfo  {journal} {Proceedings of the National Academy of Sciences}\
  }\textbf {\bibinfo {volume} {114}},\ \bibinfo {pages} {2503} (\bibinfo {year}
  {2017})}\BibitemShut {NoStop}%
\bibitem [{\citenamefont {Jackson}\ \emph {et~al.}(2007)\citenamefont
  {Jackson}, \citenamefont {Proukakis},\ and\ \citenamefont
  {Barenghi}}]{Jackson2007}%
  \BibitemOpen
  \bibfield  {author} {\bibinfo {author} {\bibfnamefont {B.}~\bibnamefont
  {Jackson}}, \bibinfo {author} {\bibfnamefont {N.~P.}\ \bibnamefont
  {Proukakis}},\ and\ \bibinfo {author} {\bibfnamefont {C.~F.}\ \bibnamefont
  {Barenghi}},\ }\href@noop {} {\bibfield  {journal} {\bibinfo  {journal}
  {Phys. Rev. A}\ }\textbf {\bibinfo {volume} {75}},\ \bibinfo {pages} {051601}
  (\bibinfo {year} {2007})}\BibitemShut {NoStop}%
\bibitem [{\citenamefont {Ohya}\ \emph {et~al.}(2019)\citenamefont {Ohya},
  \citenamefont {Watabe},\ and\ \citenamefont {Nikuni}}]{Ohya:2019uf}%
  \BibitemOpen
  \bibfield  {author} {\bibinfo {author} {\bibfnamefont {H.}~\bibnamefont
  {Ohya}}, \bibinfo {author} {\bibfnamefont {S.}~\bibnamefont {Watabe}},\ and\
  \bibinfo {author} {\bibfnamefont {T.}~\bibnamefont {Nikuni}},\ }\href@noop {}
  {\bibfield  {journal} {\bibinfo  {journal} {J. Low Temp. Phys.}\ }\textbf
  {\bibinfo {volume} {196}},\ \bibinfo {pages} {140} (\bibinfo {year}
  {2019})}\BibitemShut {NoStop}%
\bibitem [{\citenamefont {Ohya}\ \emph {et~al.}()\citenamefont {Ohya},
  \citenamefont {Watabe},\ and\ \citenamefont {Nikuni}}]{Ohi_unpublished}%
  \BibitemOpen
  \bibfield  {author} {\bibinfo {author} {\bibfnamefont {H.}~\bibnamefont
  {Ohya}}, \bibinfo {author} {\bibfnamefont {S.}~\bibnamefont {Watabe}},\ and\
  \bibinfo {author} {\bibfnamefont {T.}~\bibnamefont {Nikuni}},\ }\bibinfo
  {note} {unpublished}\BibitemShut {NoStop}%
\bibitem [{\citenamefont {Davis}\ \emph
  {et~al.}(2001{\natexlab{a}})\citenamefont {Davis}, \citenamefont {Ballagh},\
  and\ \citenamefont {Burnett}}]{Davis_2001}%
  \BibitemOpen
  \bibfield  {author} {\bibinfo {author} {\bibfnamefont {M.~J.}\ \bibnamefont
  {Davis}}, \bibinfo {author} {\bibfnamefont {R.~J.}\ \bibnamefont {Ballagh}},\
  and\ \bibinfo {author} {\bibfnamefont {K.}~\bibnamefont {Burnett}},\ }\href
  {https://doi.org/10.1088/0953-4075/34/22/316} {\bibfield  {journal} {\bibinfo
   {journal} {J. Phys. B: At. Mol. Opt. Phys.}\ }\textbf {\bibinfo {volume}
  {34}},\ \bibinfo {pages} {4487} (\bibinfo {year}
  {2001}{\natexlab{a}})}\BibitemShut {NoStop}%
\bibitem [{\citenamefont {Davis}\ \emph
  {et~al.}(2001{\natexlab{b}})\citenamefont {Davis}, \citenamefont {Morgan},\
  and\ \citenamefont {Burnett}}]{Davis2001}%
  \BibitemOpen
  \bibfield  {author} {\bibinfo {author} {\bibfnamefont {M.~J.}\ \bibnamefont
  {Davis}}, \bibinfo {author} {\bibfnamefont {S.~A.}\ \bibnamefont {Morgan}},\
  and\ \bibinfo {author} {\bibfnamefont {K.}~\bibnamefont {Burnett}},\
  }\href@noop {} {\bibfield  {journal} {\bibinfo  {journal} {Phys. Rev. Lett.}\
  }\textbf {\bibinfo {volume} {87}},\ \bibinfo {pages} {160402} (\bibinfo
  {year} {2001}{\natexlab{b}})}\BibitemShut {NoStop}%
\bibitem [{\citenamefont {Davis}\ \emph {et~al.}(2002)\citenamefont {Davis},
  \citenamefont {Morgan},\ and\ \citenamefont {Burnett}}]{Davis2002}%
  \BibitemOpen
  \bibfield  {author} {\bibinfo {author} {\bibfnamefont {M.~J.}\ \bibnamefont
  {Davis}}, \bibinfo {author} {\bibfnamefont {S.~A.}\ \bibnamefont {Morgan}},\
  and\ \bibinfo {author} {\bibfnamefont {K.}~\bibnamefont {Burnett}},\
  }\href@noop {} {\bibfield  {journal} {\bibinfo  {journal} {Phys. Rev. A}\
  }\textbf {\bibinfo {volume} {66}},\ \bibinfo {pages} {053618} (\bibinfo
  {year} {2002})}\BibitemShut {NoStop}%
\bibitem [{\citenamefont {Davis}\ and\ \citenamefont
  {Morgan}(2003)}]{Davis2003}%
  \BibitemOpen
  \bibfield  {author} {\bibinfo {author} {\bibfnamefont {M.~J.}\ \bibnamefont
  {Davis}}\ and\ \bibinfo {author} {\bibfnamefont {S.~A.}\ \bibnamefont
  {Morgan}},\ }\href@noop {} {\bibfield  {journal} {\bibinfo  {journal} {Phys.
  Rev. A}\ }\textbf {\bibinfo {volume} {68}},\ \bibinfo {pages} {053615}
  (\bibinfo {year} {2003})}\BibitemShut {NoStop}%
\bibitem [{\citenamefont {Blakie}\ and\ \citenamefont
  {Davis}(2005)}]{Blair2005}%
  \BibitemOpen
  \bibfield  {author} {\bibinfo {author} {\bibfnamefont {P.~B.}\ \bibnamefont
  {Blakie}}\ and\ \bibinfo {author} {\bibfnamefont {M.~J.}\ \bibnamefont
  {Davis}},\ }\href@noop {} {\bibfield  {journal} {\bibinfo  {journal} {Phys.
  Rev. A}\ }\textbf {\bibinfo {volume} {72}},\ \bibinfo {pages} {063608}
  (\bibinfo {year} {2005})}\BibitemShut {NoStop}%
\bibitem [{\citenamefont {Blakie}\ and\ \citenamefont
  {Davis}(2007)}]{Blakie2007}%
  \BibitemOpen
  \bibfield  {author} {\bibinfo {author} {\bibfnamefont {P.~B.}\ \bibnamefont
  {Blakie}}\ and\ \bibinfo {author} {\bibfnamefont {M.~J.}\ \bibnamefont
  {Davis}},\ }\href {https://doi.org/10.1088/0953-4075/40/11/007} {\bibfield
  {journal} {\bibinfo  {journal} {J. Phys. B: At. Mol. Opt. Phys.}\ }\textbf
  {\bibinfo {volume} {40}},\ \bibinfo {pages} {2043} (\bibinfo {year}
  {2007})}\BibitemShut {NoStop}%
\bibitem [{\citenamefont {Blakie}\ \emph {et~al.}(2008)\citenamefont {Blakie},
  \citenamefont {Bradley}, \citenamefont {Davis}, \citenamefont {Ballagh},\
  and\ \citenamefont {Gardiner}}]{Blakie2008}%
  \BibitemOpen
  \bibfield  {author} {\bibinfo {author} {\bibfnamefont {P.}~\bibnamefont
  {Blakie}}, \bibinfo {author} {\bibfnamefont {A.}~\bibnamefont {Bradley}},
  \bibinfo {author} {\bibfnamefont {M.}~\bibnamefont {Davis}}, \bibinfo
  {author} {\bibfnamefont {R.}~\bibnamefont {Ballagh}},\ and\ \bibinfo {author}
  {\bibfnamefont {C.}~\bibnamefont {Gardiner}},\ }\href@noop {} {\bibfield
  {journal} {\bibinfo  {journal} {Adv. Phys.}\ }\textbf {\bibinfo {volume}
  {57}},\ \bibinfo {pages} {363} (\bibinfo {year} {2008})}\BibitemShut
  {NoStop}%
\bibitem [{\citenamefont {Blakie}(2008)}]{Blair2008}%
  \BibitemOpen
  \bibfield  {author} {\bibinfo {author} {\bibfnamefont {P.~B.}\ \bibnamefont
  {Blakie}},\ }\href@noop {} {\bibfield  {journal} {\bibinfo  {journal} {Phys.
  Rev. E}\ }\textbf {\bibinfo {volume} {78}},\ \bibinfo {pages} {026704}
  (\bibinfo {year} {2008})}\BibitemShut {NoStop}%
\bibitem [{\citenamefont {Moulder}\ \emph {et~al.}(2012)\citenamefont
  {Moulder}, \citenamefont {Beattie}, \citenamefont {Smith}, \citenamefont
  {Tammuz},\ and\ \citenamefont {Hadzibabic}}]{Moulder2012}%
  \BibitemOpen
  \bibfield  {author} {\bibinfo {author} {\bibfnamefont {S.}~\bibnamefont
  {Moulder}}, \bibinfo {author} {\bibfnamefont {S.}~\bibnamefont {Beattie}},
  \bibinfo {author} {\bibfnamefont {R.~P.}\ \bibnamefont {Smith}}, \bibinfo
  {author} {\bibfnamefont {N.}~\bibnamefont {Tammuz}},\ and\ \bibinfo {author}
  {\bibfnamefont {Z.}~\bibnamefont {Hadzibabic}},\ }\href@noop {} {\bibfield
  {journal} {\bibinfo  {journal} {Phys. Rev. A}\ }\textbf {\bibinfo {volume}
  {86}},\ \bibinfo {pages} {013629} (\bibinfo {year} {2012})}\BibitemShut
  {NoStop}%
\bibitem [{\citenamefont {Schick}(1971)}]{PhysRevA.3.1067}%
  \BibitemOpen
  \bibfield  {author} {\bibinfo {author} {\bibfnamefont {M.}~\bibnamefont
  {Schick}},\ }\href@noop {} {\bibfield  {journal} {\bibinfo  {journal} {Phys.
  Rev. A}\ }\textbf {\bibinfo {volume} {3}},\ \bibinfo {pages} {1067} (\bibinfo
  {year} {1971})}\BibitemShut {NoStop}%
\bibitem [{\citenamefont {Bisset}\ \emph {et~al.}(2009)\citenamefont {Bisset},
  \citenamefont {Davis}, \citenamefont {Simula},\ and\ \citenamefont
  {Blakie}}]{PhysRevA.79.033626}%
  \BibitemOpen
  \bibfield  {author} {\bibinfo {author} {\bibfnamefont {R.~N.}\ \bibnamefont
  {Bisset}}, \bibinfo {author} {\bibfnamefont {M.~J.}\ \bibnamefont {Davis}},
  \bibinfo {author} {\bibfnamefont {T.~P.}\ \bibnamefont {Simula}},\ and\
  \bibinfo {author} {\bibfnamefont {P.~B.}\ \bibnamefont {Blakie}},\
  }\href@noop {} {\bibfield  {journal} {\bibinfo  {journal} {Phys. Rev. A}\
  }\textbf {\bibinfo {volume} {79}},\ \bibinfo {pages} {033626} (\bibinfo
  {year} {2009})}\BibitemShut {NoStop}%
\bibitem [{\citenamefont {Rugh}(2001)}]{Rugh2001}%
  \BibitemOpen
  \bibfield  {author} {\bibinfo {author} {\bibfnamefont {H.~H.}\ \bibnamefont
  {Rugh}},\ }\href@noop {} {\bibfield  {journal} {\bibinfo  {journal} {Phys.
  Rev. E}\ }\textbf {\bibinfo {volume} {64}},\ \bibinfo {pages} {055101}
  (\bibinfo {year} {2001})}\BibitemShut {NoStop}%
\bibitem [{\citenamefont {Gawryluk}\ and\ \citenamefont
  {Brewczyk}(2021)}]{Gawryluk:2021wa}%
  \BibitemOpen
  \bibfield  {author} {\bibinfo {author} {\bibfnamefont {K.}~\bibnamefont
  {Gawryluk}}\ and\ \bibinfo {author} {\bibfnamefont {M.}~\bibnamefont
  {Brewczyk}},\ }\href@noop {} {\bibfield  {journal} {\bibinfo  {journal} {Sci
  Rep}\ }\textbf {\bibinfo {volume} {11}},\ \bibinfo {pages} {10773} (\bibinfo
  {year} {2021})}\BibitemShut {NoStop}%
\bibitem [{\citenamefont {Tsuzuki}(1971)}]{Tsuzuki:1971vw}%
  \BibitemOpen
  \bibfield  {author} {\bibinfo {author} {\bibfnamefont {T.}~\bibnamefont
  {Tsuzuki}},\ }\href@noop {} {\bibfield  {journal} {\bibinfo  {journal}
  {Journal of Low Temperature Physics}\ }\textbf {\bibinfo {volume} {4}},\
  \bibinfo {pages} {441} (\bibinfo {year} {1971})}\BibitemShut {NoStop}%
\bibitem [{\citenamefont {Metz}\ \emph {et~al.}(2021)\citenamefont {Metz},
  \citenamefont {Polo}, \citenamefont {Weber},\ and\ \citenamefont
  {Busch}}]{FriederikeMetz:2021bh}%
  \BibitemOpen
  \bibfield  {author} {\bibinfo {author} {\bibfnamefont {F.}~\bibnamefont
  {Metz}}, \bibinfo {author} {\bibfnamefont {J.}~\bibnamefont {Polo}}, \bibinfo
  {author} {\bibfnamefont {N.}~\bibnamefont {Weber}},\ and\ \bibinfo {author}
  {\bibfnamefont {T.}~\bibnamefont {Busch}},\ }\href@noop {} {\bibfield
  {journal} {\bibinfo  {journal} {Mach. Learn.: Sci. Technol.}\ }\textbf
  {\bibinfo {volume} {2}},\ \bibinfo {pages} {035019} (\bibinfo {year}
  {2021})}\BibitemShut {NoStop}%
\bibitem [{\citenamefont {Guo}\ \emph {et~al.}(2021)\citenamefont {Guo},
  \citenamefont {Fritsch}, \citenamefont {Greenberg}, \citenamefont
  {Spielman},\ and\ \citenamefont {Zwolak}}]{ShangjieGuo:2021iy}%
  \BibitemOpen
  \bibfield  {author} {\bibinfo {author} {\bibfnamefont {S.}~\bibnamefont
  {Guo}}, \bibinfo {author} {\bibfnamefont {A.~R.}\ \bibnamefont {Fritsch}},
  \bibinfo {author} {\bibfnamefont {C.}~\bibnamefont {Greenberg}}, \bibinfo
  {author} {\bibfnamefont {I.~B.}\ \bibnamefont {Spielman}},\ and\ \bibinfo
  {author} {\bibfnamefont {J.~P.}\ \bibnamefont {Zwolak}},\ }\href@noop {}
  {\bibfield  {journal} {\bibinfo  {journal} {Mach. Learn.: Sci. Technol.}\
  }\textbf {\bibinfo {volume} {2}},\ \bibinfo {pages} {035020} (\bibinfo {year}
  {2021})}\BibitemShut {NoStop}%
\bibitem [{\citenamefont {Yang}(1962)}]{Yang1962}%
  \BibitemOpen
  \bibfield  {author} {\bibinfo {author} {\bibfnamefont {C.~N.}\ \bibnamefont
  {Yang}},\ }\href@noop {} {\bibfield  {journal} {\bibinfo  {journal} {Rev.
  Mod. Phys.}\ }\textbf {\bibinfo {volume} {34}},\ \bibinfo {pages} {694}
  (\bibinfo {year} {1962})}\BibitemShut {NoStop}%
\bibitem [{\citenamefont {Penrose}\ and\ \citenamefont
  {Onsager}(1956)}]{Penrose1956}%
  \BibitemOpen
  \bibfield  {author} {\bibinfo {author} {\bibfnamefont {O.}~\bibnamefont
  {Penrose}}\ and\ \bibinfo {author} {\bibfnamefont {L.}~\bibnamefont
  {Onsager}},\ }\href@noop {} {\bibfield  {journal} {\bibinfo  {journal} {Phys.
  Rev.}\ }\textbf {\bibinfo {volume} {104}},\ \bibinfo {pages} {576} (\bibinfo
  {year} {1956})}\BibitemShut {NoStop}%
\end{thebibliography}%

\end{document}